\documentclass[11pt,draftclsnofoot,onecolumn]{IEEEtran}
\usepackage{amsmath}
\usepackage{amssymb}
\usepackage{amsthm}
\usepackage{graphicx}
\usepackage{cite}
\usepackage{subfigure}
\usepackage{algorithm,algorithmic}
\theoremstyle{remark}
\newtheorem{rem}{Remark}

\newcommand{\mat}[1]{#1}

\newcommand{\Tr}[1]{{\mathrm{Tr}}\left\{#1\right\}}

\newcommand{\range}[1]{\mathcal{R}\left(#1\right)}
\newcommand{\etr}[1]{{\mathrm{etr}}\left\{#1\right\}}
\newcommand{\diag}[1]{{\mathrm{diag}}\left(#1\right)}

\newcommand{\A}{\mat{A}}

\renewcommand{\H}{\mat{H}}
\newcommand{\Hhat}{\hat{\H}}

\newcommand{\I}{\mat{I}}

\newcommand{\N}{\mat{N}}
\newcommand{\Q}{\mat{Q}}
\newcommand{\mS}{\mat{S}}

\newcommand{\X}{\mat{X}}
\newcommand{\Y}{\mat{Y}}
\newcommand{\Z}{\mat{Z}}

\newcommand{\varn}{\sigma^{2}}
\newcommand{\ivarn}{\sigma^{-2}}
\newcommand{\kbar}{\bar{k}}

\newcommand{\Nbi}{N_{\text{bi}}}
\newcommand{\Nr}{N_{\text{r}}}
\newcommand{\Nit}{N_{\text{it}}}

\newcommand{\pdfN}[2]{\mathcal{N}\left( #1,#2\right)}
\newcommand{\Bgh}[1]{\mathrm{B}\left( #1\right)} 

\begin{document}
\title{Joint Bayesian Estimation of Close Subspaces from Noisy Measurements}

\author{\vspace{1cm}Olivier Besson$^1$, Nicolas Dobigeon$^2$ and
Jean-Yves Tourneret$^2$
\\
\normalsize $^1$ University of Toulouse, ISAE, Dept. Electronics Optronics Signal, 31055 Toulouse, France \\
\small\texttt{Olivier.Besson@isae.fr}\\
\normalsize $^2$ University of Toulouse, IRIT/INP-ENSEEIHT/T\'eSA, 31071 Toulouse, France. \\
\small\texttt{\{Nicolas.Dobigeon,Jean-Yves.Tourneret\}@enseeiht.fr}}

\maketitle
\begin{abstract}
In this letter, we consider two sets of observations defined as subspace signals embedded in noise and we wish to analyze the distance between these two subspaces. The latter entails evaluating the angles between the subspaces, an issue reminiscent of the well-known Procrustes problem. A Bayesian approach is investigated where the subspaces of interest are considered as random with a joint prior distribution (namely a Bingham distribution), which allows the closeness of the two subspaces to be adjusted. Within this framework, the minimum mean-square distance estimator of both subspaces is formulated and implemented via a Gibbs sampler. A simpler scheme based on alternative maximum a posteriori estimation is also presented. The new schemes are shown to provide more accurate estimates of the angles between the subspaces, compared to singular value decomposition based independent estimation of the two subspaces.
\end{abstract}

\newpage
\section{Problem statement}
Modeling signals of interest as belonging to a linear subspace is arguably one of the most encountered approach in engineering applications \cite{Scharf91,Scharf91b,VanderVeen93}. Estimation of such signals in additive white noise is usually conducted via the singular value decomposition which has proven to be very successful in numerous problems, including spectral analysis or direction finding. In this letter, we consider a situation where two independent noisy observations of a subspace signal are available but, due to miscalibration or a change in the observed process, the subspace of interest is slightly different from one observation to the other. More precisely, assume that we observe two $M \times T$ matrices $\X_{1}$ and $\X_{2}$ given by
\begin{equation}
\X_{k} = \H_{k} \mS_{k} + \N_{k}; \quad k=1,2
\end{equation}
where the orthogonal $M \times R$ matrices $\H_{k}$ ($\H_{k}^{T} \H_{k}=\I_{R}$) span the subspace where the signals of interest lie, $\mS_{k}$ stands for the  matrix of coordinates of the noise-free data within the range space $\range{\H_{k}}$ of $\H_{k}$, and $\N_{k}$ denotes an additive white Gaussian noise. Herein, we are interested in recovering the subspaces $\H_{1}$, $\H_{2}$ but, maybe more importantly, to have an indication of the ``difference'' between these two subspaces. The natural distance between  $\H_{1}$ and $\H_{2}$ is given by $\left[ \sum_{r=1}^{R} \theta_{r}^{2} \right]^{1/2}$ where $\theta_{r}$ are the principal angles between $\H_{1}$ and $\H_{2}$, which can be obtained from the singular value decomposition (SVD) $\H_{1}^{T} \H_{2} = \Y \diag{\cos \theta_{1}, \ldots, \cos \theta_{R}} \Z^{T}$. This problem is somehow reminiscent of the orthogonal matrix Procrustes problem \cite[p. 601]{Golub96} where one seeks an orthogonal matrix that brings $\H_{1}$ close to $\H_{2}$ by solving $\min_{\Q^{T}\Q=\I} \left\| \H_{2} - \H_{1} \Q \right\|_{F}$.
The solution is well known to be $\Q = \Y \Z^{T}$. The problem here is slightly different as we only have access to $\X_{1}$, $\X_{2}$ and not to the subspaces themselves. Moreover, we would like to exploit the fact that $\H_{1}$ and $\H_{2}$ are close subspaces. In order to embed this knowledge, a Bayesian framework is formulated where $\H_{1}$ and $\H_{2}$ are treated as random matrices with a joint distribution, as detailed now.

Let us state our assumptions and our approach to estimating $\H_{1}$, $\H_{2}$ and subsequently the principal angles $\theta_{r}$, $r=1,\cdots,R$. Assuming that the columns of $\N_{1}$ and $\N_{2}$ are independent and identically Gaussian distributed $\N_{k} \sim \pdfN{\mat{0}}{\varn \I}$ with $\varn$ known, the likelihood function of $\X_{k}$ is given by
\begin{equation}
p(\X_{k} | \H_{k} , \mS_{k} ) \propto \etr{-\frac{1}{2\varn} \left(\X_{k} - \H_{k} \mS_{k} \right)^{T} \left(\X_{k} - \H_{k} \mS_{k} \right)}
\end{equation}
where $\propto$ means proportional to and $\etr{.}$ stands for the exponential of the trace of the matrix between braces. As for $\mS_{k}$, we assume that no knowledge about it is available so that its prior distribution is given by $\pi(\mS_{k}) \propto 1$. Note that this is an improper prior but, as will be shown shortly, marginalizing with respect to $\mS_{k}$ results in a proper distribution. Indeed,
\begin{align}\label{p(Xk|Hk)}
p(\X_{k} | \H_{k}) & = \int p(\X_{k} | \H_{k} , \mS_{k} ) \pi(\mS_{k}) d \mS_{k} \nonumber \\
& \propto \etr{-\frac{1}{2\varn} \left( \X_{k}^{T} \X_{k} - \X_{k}^{T} \H_{k} \H_{k}^{T} \X_{k} \right) }.
\end{align}
Let us now turn to our assumption regarding $\H_{1}$ and $\H_{2}$. We assume that $\H_{1}$ is uniformly distributed on the Stiefel manifold \cite{Chikuse03} and that $\H_{2}$, conditioned on $\H_{1}$, follows a Bingham distribution \cite{Chikuse03,Mardia99} with parameter matrix $\kappa \H_{1} \H_{1}^{T}$, i.e.,
\begin{equation}\label{pi(H2|H1)}
\pi(\H_{2} | \H_{1} ) = C(\lambda(\H_{1})) \etr{ \kappa \H_{2}^{T} \H_{1} \H_{1}^{T} \H_{2}}
\end{equation}
where $C(\lambda(\H_{1}))$ is a constant that depends on the eigenvalues of $\H_{1}$. The scalar parameter $\kappa$ directly impacts the prior distribution of the angles between $\range{\H_{1}}$ and $\range{\H_{2}}$ and therefore its value should reflect our knowledge about the closeness between these two subspaces. Briefly stated, the larger $\kappa$ the closer $\range{\H_{1}}$ and $\range{\H_{2}}$.

\section{Subspace estimation}
Our objective is, given the likelihood function in \eqref{p(Xk|Hk)} and the prior in \eqref{pi(H2|H1)}, to estimate $\H_{1}$,  $\H_{2}$ and then deduce the principal angles between them. Towards this end, let us first write the joint posterior distribution of $\H_{1}$ and $\H_{2}$ as
\begin{align}\label{p(H1,H2|X1,X2)}
&p(\H_{1},\H_{2} | \X_{1},\X_{2}) \propto p(\X_{1},\X_{2} | \H_{1},\H_{2}) \pi(\H_{2} | \H_{1} )  \pi( \H_{1} ) \nonumber \\
&\propto \etr{ \frac{1}{2\varn} \X_{1}^{T} \H_{1} \H_{1}^{T} \X_{1} + \frac{1}{2\varn} \X_{2}^{T} \H_{2} \H_{2}^{T} \X_{2} } \nonumber \\
& \times  \etr{ \kappa \H_{2}^{T} \H_{1} \H_{1}^{T} \H_{2}}.
\end{align}
In the sequel we let $\kbar = \left\{1,2\right\} \setminus k$.
The posterior density of $\H_{k}$ only is thus
\begin{align}\label{p(Hk|X1,X2)}
&p(\H_{k} |   \X_{1} ,\X_{2}) = \int p(\H_{k},\H_{\kbar} | \X_{1},\X_{2}) d \H_{\kbar}  \nonumber \\
& \propto  \etr{ \frac{1}{2\varn} \X_{k}^{T} \H_{k} \H_{k}^{T} \X_{k}} \nonumber \\
&\times \int \etr{ \H_{\kbar}^{T} \left[ \frac{1}{2\varn} \X_{\kbar} \X_{\kbar}^{T}  + \kappa \H_{k} \H_{k}^{T} \right]  \H_{\kbar}} d\H_{\kbar} \nonumber \\
&\propto  C(\lambda(\frac{1}{2\varn} \X_{\kbar} \X_{\kbar}^{T}  + \kappa \H_{k} \H_{k}^{T})) \etr{ \frac{1}{2\varn} \X_{k}^{T} \H_{k} \H_{k}^{T} \X_{k}}.
\end{align}
The minimum mean-square distance (MMSD) estimator of $\H_{k}$ is defined as \cite{Besson11}
\begin{equation}\label{MMSD(Hk)}
\Hhat_{k-\text{\tiny{MMSD}}} = \mathcal{P}_{R}\left\{ \int \H_{k} \H_{k}^T p(\H_{k} | \X_{1}, \X_{2} ) d \H_{k} \right\}
\end{equation}
where $\mathcal{P}_{R}\left\{\cdot \right\}$ stands for the $R$ principal eigenvectors of the matrix between braces. From inspection of  $p(\H_{k} |   \X_{1} ,\X_{2})$, the above integral in \eqref{MMSD(Hk)} does not seem to be tractable. Therefore, we turn to Markov chain Monte-Carlo (MCMC) simulation methods to approximate it \cite{Robert04}. The first idea that comes to mind is to generate samples drawn  from $p(\H_{k} | \X_{1}, \X_{2} )$ and to approximate the integral by an arithmetic mean. However, the distribution in \eqref{p(Hk|X1,X2)} is not obvious to sample. On the contrary, the conditional distribution of $\H_{k} | \H_{\kbar} ,\X_{1},\X_{2}$ belongs to a known family. Indeed,  from \eqref{p(H1,H2|X1,X2)} one has
\begin{equation}\label{post_cond_p}
p(\H_{k} |\H_{\kbar} , \X_{1},\X_{2}) \propto \etr{ \H_{k}^{T} \left[ \frac{1}{2\varn} \X_{k} \X_{k}^{T}  + \kappa \H_{\kbar} \H_{\kbar}^{T} \right] \H_{k} }
\end{equation}
which is recognized as a Bingham distribution, i.e.,
\begin{equation}\label{post_cond_distrib}
\H_{k} |\H_{\kbar} , \X_{1} ,  \X_{2} \sim  \Bgh{\frac{1}{2\varn} \X_{k} \X_{k}^{T}  + \kappa \H_{\kbar} \H_{\kbar}^{T}} .
\end{equation}
This leads us to consider a Gibbs sampling scheme which, uses \eqref{post_cond_distrib} to draw samples asymptotically distributed according to  $p(\H_{k} |   \X_{1} ,\X_{2})$. This scheme is summarized in Table \ref{algo:gibbs}.
\begin{table}[h]
\caption{Gibbs sampler for estimation of $\H_{1}$ and $\H_{2}$.}
\begin{algorithmic}[1]
\REQUIRE initial value $\H_{1}(0)$
\FOR{$n=1,\cdots,\Nbi + \Nr$}
\STATE sample $\H_{2}(n)$  from $\Bgh{\frac{1}{2\varn} \X_{2} \X_{2}^{T}  + \kappa \H_{1}(n-1) \H_{1}(n-1)^{T} }$.
\STATE sample $\H_{1}(n)$ from  $\Bgh{\frac{1}{2\varn} \X_{1} \X_{1}^{T}  + \kappa \H_{2}(n) \H_{2}(n)^{T} }$.
\ENDFOR
\ENSURE sequence of random matrices $\H_{1}(n)$ and $\H_{2}(n)$.
\end{algorithmic}
\label{algo:gibbs}
\end{table}

Once a set of $\Nr$ matrices $\H_{1}(n)$ and $\H_{2}(n)$ has been generated, the MMSD estimator of $\H_{k}$ can be approximated as
\begin{equation}\label{Hk_MMSD}
\Hhat_{k-\text{\tiny{MMSD}}} = \mathcal{P}_{R}\left\{ \Nr^{-1} \sum_{n=\Nbi+1}^{\Nbi + \Nr} \H_{k}(n) \H_{k}(n)^{T} \right\}.
\end{equation}

An alternative and possibly more computationally efficient approach would entail considering maximum a posteriori (MAP) estimation. However, the joint MAP estimation of $\H_{1}$ and $\H_{2}$ from $p(\H_{1},\H_{2} | \X_{1},\X_{2})$ in \eqref{p(H1,H2|X1,X2)} does not appear tractable. It is in fact customary in this case to consider iterative alternate maximization of $p(\H_{1},\H_{2} | \X_{1},\X_{2})$, i..e, maximize it first with respect to $\H_{1}$  holding $\H_{2}$ fixed, and then with respect to $\H_{2}$  holding $\H_{1}$ fixed. This method guarantees that $p(\H_{1},\H_{2} | \X_{1},\X_{2})$ increases along the iterations. Moreover, at each step, the MAP estimation of one matrix, conditioned on the other one, is simple as
\begin{align}\label{Hkmap|Hkbar}
\Hhat_{k-\text{\tiny{MAP}}} | \H_{\kbar} &= \arg \max_{\H_{k}} p(\H_{k} |\H_{\kbar} , \X_{1},\X_{2}) \nonumber \\
&=  \mathcal{P}_{R}\left\{  \frac{1}{2\varn} \X_{k} \X_{k}^{T}  + \kappa \H_{\kbar} \H_{\kbar}^{T} \right\}.
\end{align}
Note that \eqref{Hkmap|Hkbar} is also the MMSD estimator of $\H_{k}$ given $\H_{\kbar}$ since, if $\H \sim \Bgh{\A}$,  the MMSD estimator of $\H$ is simply  $\mathcal{P}_{R}\left\{ \A \right\}$ \cite{Besson11}.  Therefore we propose the scheme of Table \ref{algo:imap} which we refer to as iterative MAP (iMAP). This approach  may be more computationally efficient than the Gibbs sampler, particularly if $\Nit \ll \Nr$.
\begin{table}[h]
\caption{Iterative MAP estimation of $\H_{1}$ and $\H_{2}$.}
\begin{algorithmic}[1]
\REQUIRE initial value $\H_{1}(0)$
\FOR{$n=1,\cdots,\Nit$}
\STATE evaluate  $\H_{2}(n) = \mathcal{P}_{R}\left\{  \frac{1}{2\varn} \X_{2} \X_{2}^{T}  + \kappa \H_{1}(n-1) \H_{1}(n-1)^{T} \right\}$.
\STATE evaluate  $\H_{1}(n) = \mathcal{P}_{R}\left\{  \frac{1}{2\varn} \X_{1} \X_{1}^{T}  + \kappa \H_{2}(n) \H_{2}(n)^{T} \right\}$.
\ENDFOR
\ENSURE  $\Hhat_{k-\text{\tiny{MAP}}} = \H_{k}(\Nit)$.
\end{algorithmic}
\label{algo:imap}
\end{table}

\begin{rem} (\emph{estimation by regularization})
We have decided in this work to embed the knowledge that $\range{\H_{1}}$ is close to $\range{\H_{2}}$ in a prior distribution. An alternative would be to consider regularized maximum likelihood estimation (MLE). More precisely, one may wish to maximize the likelihood function under the constraint that  $\range{\H_{1}}$ is close to $\range{\H_{2}}$. Such an approach would amount to consider the following optimization problem:
\begin{align}
\min_{\H_{1},\H_{2},\mS_{1},\mS_{2}} &- \log p(\X_{1},\X_{2} | \H_{1},\H_{2},\mS_{1},\mS_{2}) \nonumber \\
&+ \mu \left\| \H_{1}\H_{1}^{T} - \H_{2}\H_{2}^{T} \right\|_{F}^{2}.
\end{align}
Solving for $\mS_{1},\mS_{2}$ and concentrating the criterion, one ends up with minimizing
\begin{align}
&J(\H_{1},\H_{2}) = \Tr{ \frac{1}{2\varn} \X_{1}^{T} \H_{1} \H_{1}^{T} \X_{1}} \nonumber \\
& + \Tr{ \frac{1}{2\varn} \X_{2}^{T} \H_{2} \H_{2}^{T} \X_{2}} + \Tr{ 2 \mu \H_{2}^{T} \H_{1} \H_{1}^{T} \H_{2}}.
\end{align}
From observation of \eqref{p(H1,H2|X1,X2)} this is tantamount to maximizing $p(\H_{1},\H_{2} | \X_{1},\X_{2})$ with the regularization parameter $2\mu$ playing a similar role as $\kappa$. However, there are two differences. First, in a Bayesian setting $\kappa$ can be fixed by looking at the prior distribution of the angles between $\range{\H_{1}}$ and $\range{\H_{2}}$ and making it match our prior knowledge. Second, the Bayesian framework enables one to consider an MMSD estimator while the frequentist approach bears much resemblance with a maximum a posteriori estimator.
\end{rem}

\begin{rem} (\emph{alternative prior modeling})
Instead of considering a Bingham distribution as prior for $\pi(\H_{2} | \H_{1} )$ a von Mises-Fisher (vMF) distribution \cite{Mardia99} defined as
\begin{equation}
\pi(\H_{2} | \H_{1} ) \propto \etr{ c \H_{2}^{T} \H_{1}}
\end{equation}
might have been used. Under this hypothesis, it is straightforward to show that the conditional posterior distribution $p(\H_{k} |\H_{\kbar} , \X_{1},\X_{2})$ is now Bingham von Mises-Fisher (BMF)
\begin{equation}
p(\H_{k} |\H_{\kbar} , \X_{1},\X_{2}) \propto  \etr{ \frac{1}{2\varn}  \H_{k}^{T} \X_{k} \X_{k}^{T} \H_{k}  + c \H_{k}^{T}  \H_{\kbar} }  .
\end{equation}
The Gibbs sampling scheme needs to be adapted to these new distributions. However, for a BMF distribution, there does not exist a closed-form expression for the MAP estimator which means that the iterative scheme of Algorithm \ref{algo:imap} cannot be extended.
\end{rem}

\begin{rem}(\emph{Extension to more than 2 subspaces}) Let us consider a situation where $K > 2$ data matrices $\X_{k} = \H_{k} \mS_{k} + \N_{k}$ are available, so that their joint distribution, conditioned on $\H_{1 \cdots K}$ can be written as
\begin{equation}
p(\X_{1 \cdots K} | \H_{1 \cdots K})  \propto \etr{-\frac{1}{2\varn} \sum_{k=1}^{K} \left( \X_{k}^{T} \X_{k} - \X_{k}^{T} \H_{k} \H_{k}^{T} \X_{k} \right) }.
\end{equation}
Let us still assume that $\H_{1}$ is uniformly distributed on the Stiefel manifold and that $\H_{k}$ ($k > 2$), conditioned on $\H_{k-1}$, follows a Bingham distribution with parameter matrix $\kappa_{k} \H_{k-1} \H_{k-1}^{T}$, i.e.,
\begin{equation}\label{pi(Hk|Hk-1)}
\pi(\H_{k} | \H_{k-1} ) \propto  \etr{ \kappa_{k} \H_{k}^{T} \H_{k-1} \H_{k-1}^{T} \H_{k}}.
\end{equation}
Then the joint posterior distribution of $\H_{1 \cdots K}$ writes
\begin{align}
&p(\H_{1 \cdots K} | \X_{1 \cdots K} )  \propto \etr{\frac{1}{2\varn} \sum_{k=1}^{K}  \X_{k}^{T} \H_{k} \H_{k}^{T} \X_{k} } \\
&\times \etr{ \sum_{k=2}^{K} \kappa_{k} \H_{k}^{T} \H_{k-1} \H_{k-1}^{T} \H_{k}} .
\end{align}
It ensues that the conditional posterior distribution of $\H_{k}$ is given by
\begin{subequations}
\begin{align}
\H_{1} |\H_{2 \cdots K } , \X_{1 \cdots K} & \sim  \Bgh{\frac{1}{2\varn} \X_{1} \X_{1}^{T}  + \kappa_{2} \H_{2} \H_{2}^{T}} \\
\H_{k} |\H_{-k} , \X_{1 \cdots K} & \sim  \Bgh{\frac{1}{2\varn} \X_{k} \X_{k}^{T}  + \kappa_{k} \H_{k-1} \H_{k-1}^{T} }.
\end{align}
\end{subequations}
The Gibbs sampling scheme of Table \ref{algo:gibbs} as well as the iterative MAP algorithm of Table \ref{algo:imap} can be straightforwardly modified so as to account for this more general setting.
\end{rem}

\section{Numerical illustrations}
Let us now give some illustrative examples about the estimators developed above. We consider a scenario with $M=8$ and $R=2$. The two algorithms described above (referred to as GS and iMAP in the figures, respectively) will be compared to a conventional SVD-based approach where $\H_{k}$ is estimated from the $R$ dominant left singular vectors of the data matrix $\X_{k}$. For each algorithm, the angles between $\H_{1}$ and $\H_{2}$ will be estimated from the singular value decomposition of $\Hhat_{1}^{T} \Hhat_{2}$, where $\Hhat_{1},\Hhat_{2}$ stand for one of the three estimates mentioned previously. Two criteria will be used to assess the performance of the estimators. First, the MSD between $\Hhat_{k}$ and $\H_{k}$ will be used: this gives an idea of how accurately each subspace individually  is estimated. Next, since the difference between $\H_{1}$ and $\H_{2}$ is of utmost importance, we will also pay attention to the mean and standard deviation of $\hat{\theta}_{r}$ as these angles characterize how $\H_{2}$ has been moved apart from $\H_{1}$.

In all simulations the entries of $\mS_{1}$ and $\mS_{2}$ were generated as i.i.d. $\pdfN{0}{1}$. The subspaces $\H_{1}$ and $\H_{2}$ were fixed and the \emph{true} angles between them are equal to $10^{\circ}$ and $25^{\circ}$ respectively. Note that the subspaces $\H_{1}$ and $\H_{2}$ are not generated according to the prior distributions assumed above. The signal to noise ratio (SNR) is defined as $\mathrm{SNR}=\ivarn M^{-1}R$. For the Bayesian estimators, we set $\Nbi=10$, $\Nr=200$ and $\Nit=50$. In Figure \ref{fig: res_vs_T_M=8_R=2_kappa=40_SNR=0} we plot the performance versus $T$, for $\kappa=40$, while Figure \ref{fig: res_vs_SNR_M=8_R=2_kappa=40_T=6} studies the performance versus SNR. The following observations can be made:
\begin{itemize}
\item The Bayesian estimates of the individual subspaces outperform the SVD-based estimates, especially for a small number of snapshots  or a low SNR. When SNR increases however, the SVD-based estimates produce accurate estimates of each subspace.
\item The SVD-based estimator does not accurately estimate the angles between $\H_{1}$ and $\H_{2}$, unless SNR is large. In contrast, the Bayesian estimators provide a rather accurate estimation of $\theta_{r}$.
\item The Gibbs sampler is seen to perform better that the iMAP estimator, at the price of a larger computational cost however.
\end{itemize}

\ifCLASSOPTIONdraftcls
\begin{figure}[p]
\centering
\subfigure[][]{\label{fig: dist1_vs_T_M=8_R=2_kappa=40_SNR=0}
\includegraphics[width=7cm]{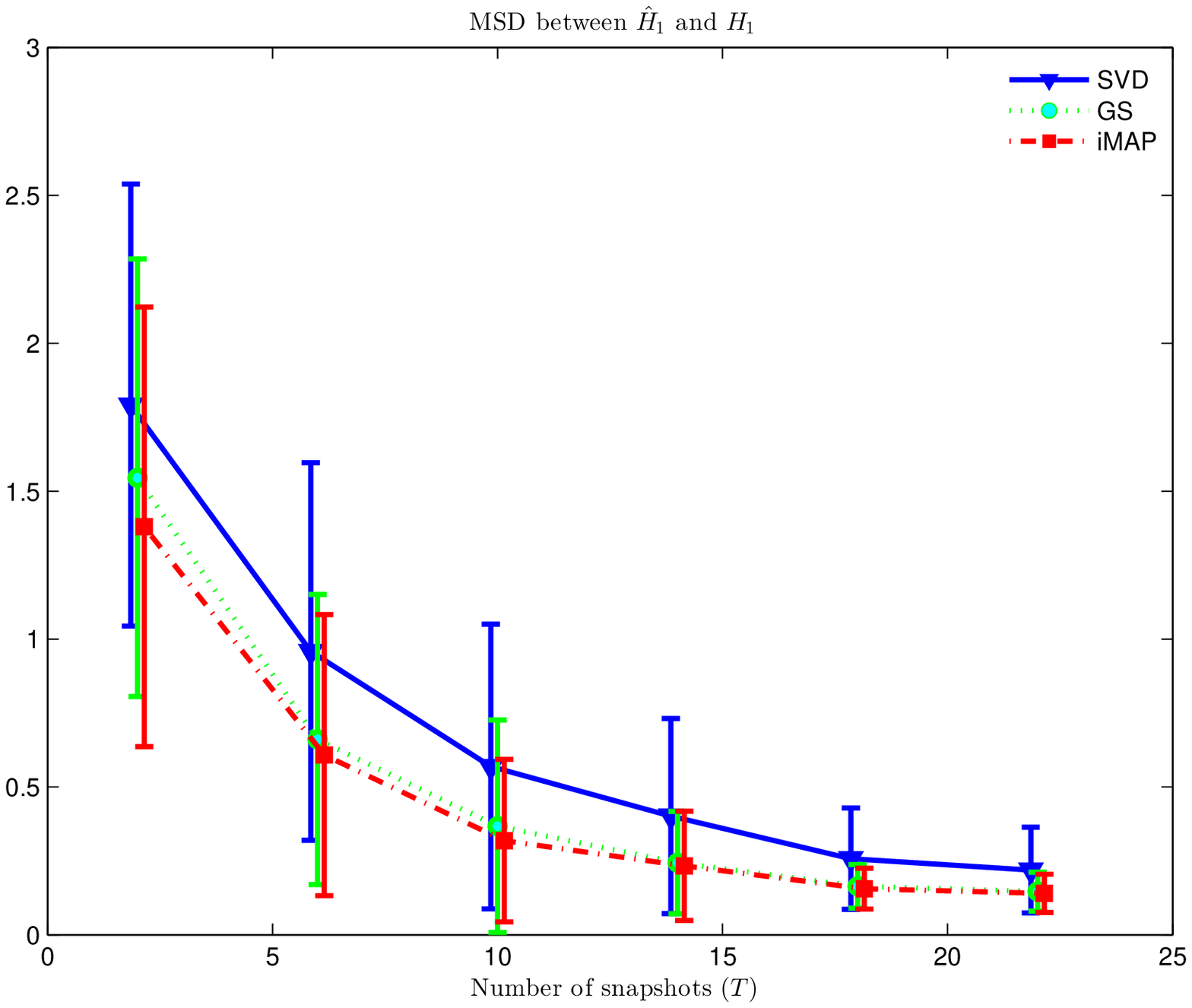}}
\subfigure[][]{\label{fig: dist2_vs_T_M=8_R=2_kappa=40_SNR=0}
\includegraphics[width=7cm]{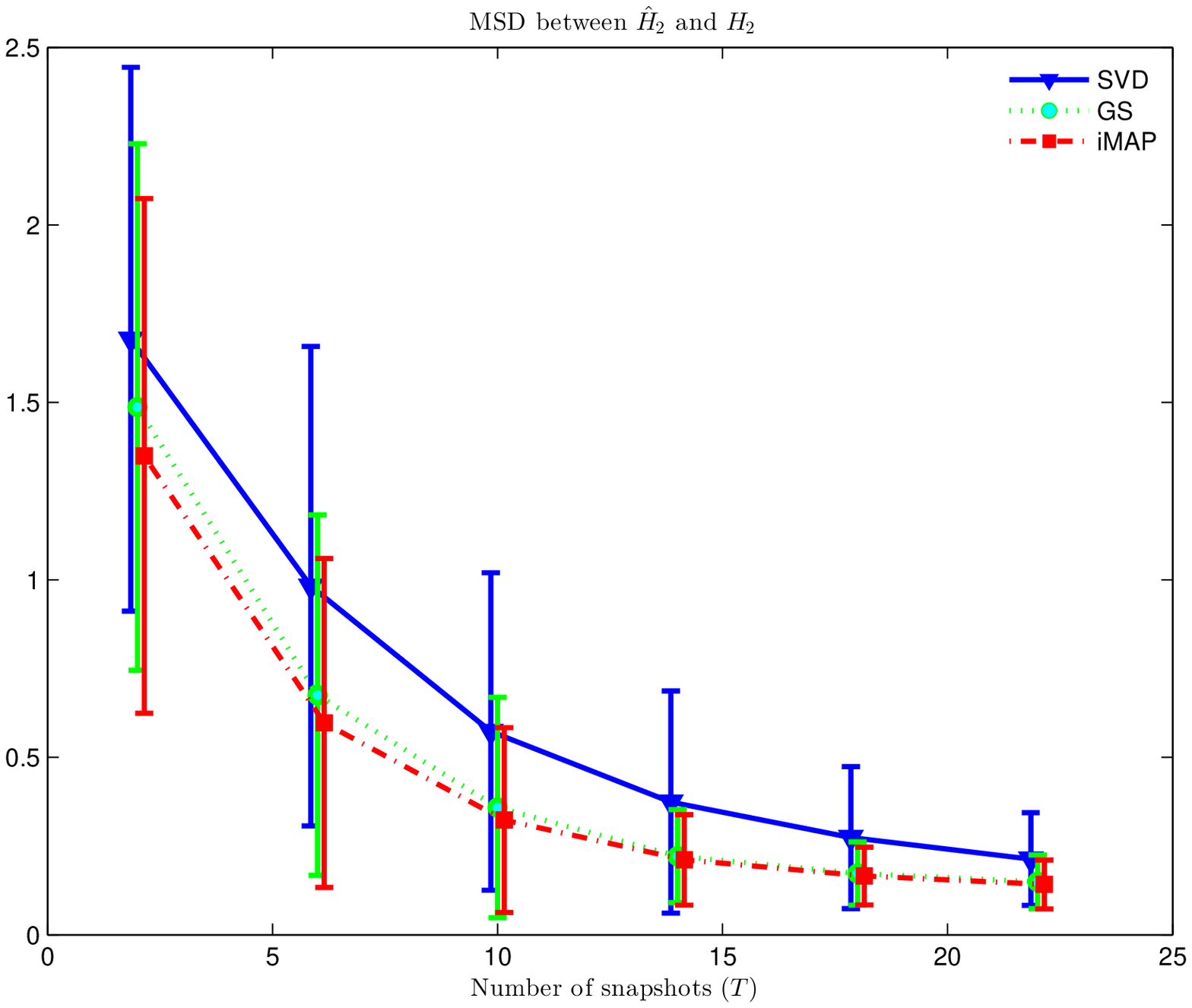}} \\
\subfigure[][]{\label{fig: theta1_vs_T_M=8_R=2_kappa=40_SNR=0}
\includegraphics[width=7cm]{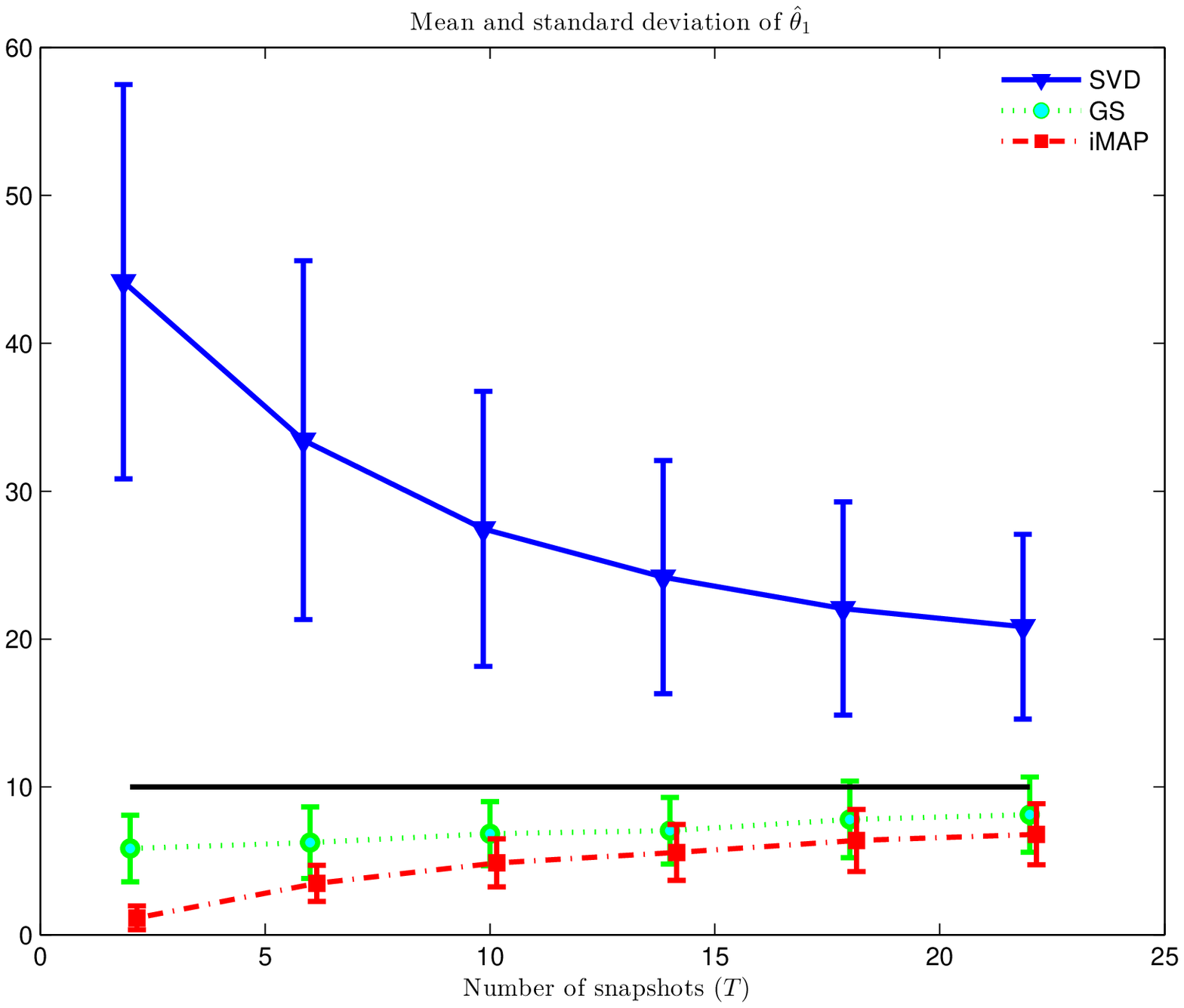}}
\subfigure[][]{\label{fig: theta2_vs_T_M=8_R=2_kappa=40_SNR=0}
\includegraphics[width=7cm]{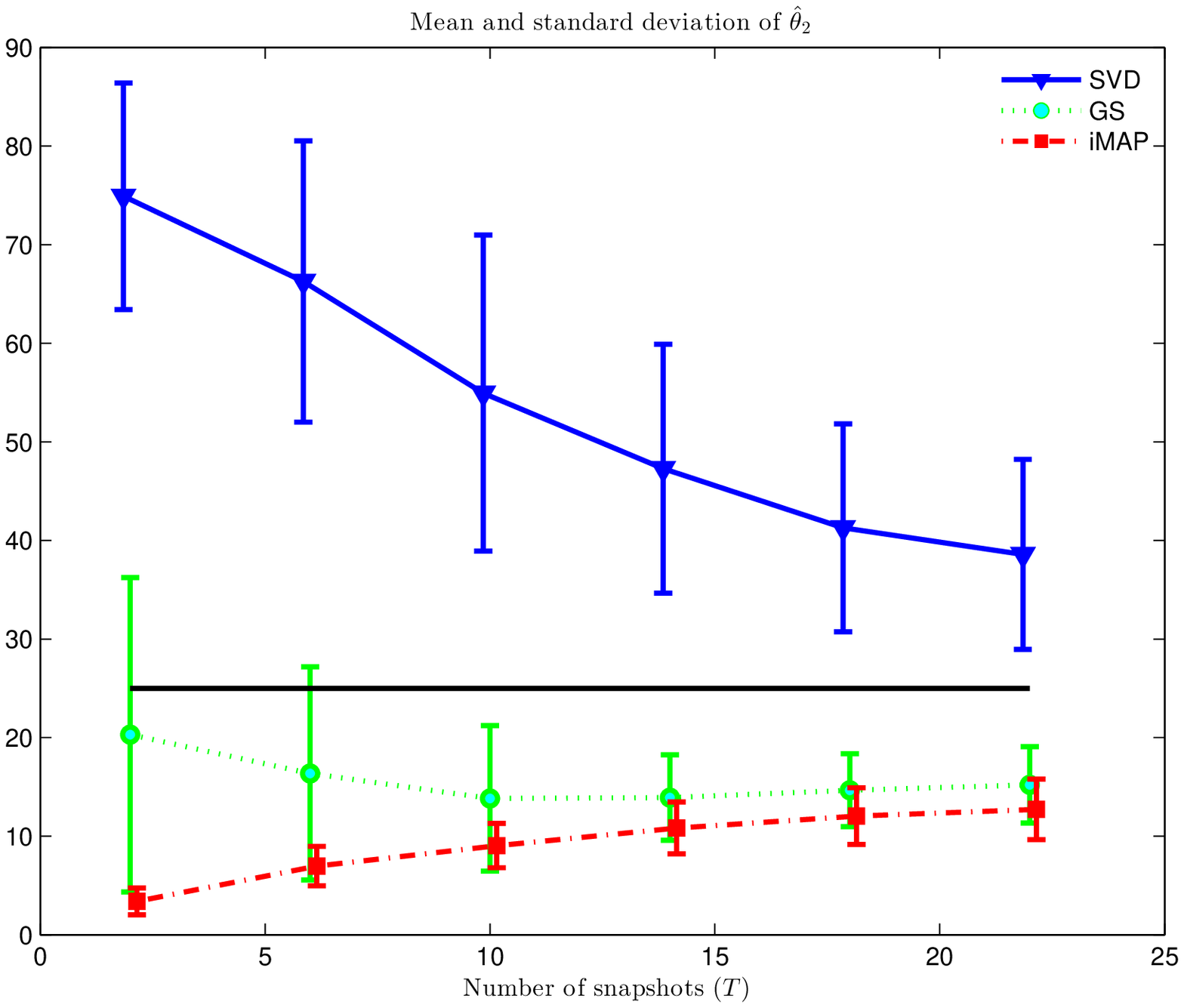}}
\caption{Performance of the estimators versus $T$. $\kappa=40$ and $\mathrm{SNR}=0$dB. \subref{fig: dist1_vs_T_M=8_R=2_kappa=40_SNR=0} $MSD(\Hhat_{1},\H_{1})$, \subref{fig: dist2_vs_T_M=8_R=2_kappa=40_SNR=0} $MSD(\Hhat_{2},\H_{2})$, \subref{fig: theta1_vs_T_M=8_R=2_kappa=40_SNR=0}, mean and std of $\hat{\theta}_{1}$,  \subref{fig: theta2_vs_T_M=8_R=2_kappa=40_SNR=0}, mean and std of $\hat{\theta}_{2}$.}
\label{fig: res_vs_T_M=8_R=2_kappa=40_SNR=0}
\end{figure}
\else
\begin{figure}[!t]
\centering
\subfigure[][]{\label{fig: dist1_vs_T_M=8_R=2_kappa=40_SNR=0}
\includegraphics[width=6cm]{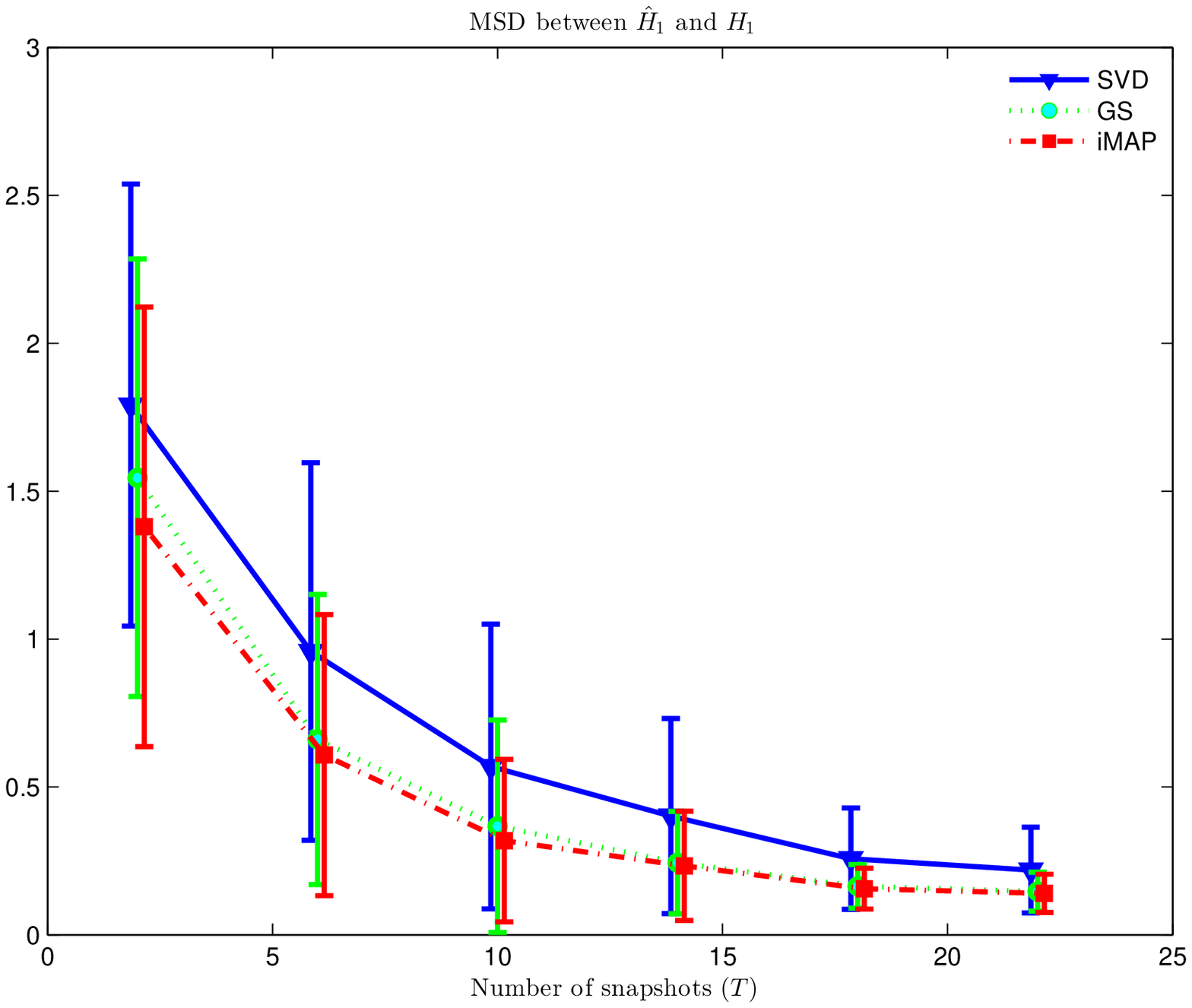}} \\
\subfigure[][]{\label{fig: dist2_vs_T_M=8_R=2_kappa=40_SNR=0}
\includegraphics[width=6cm]{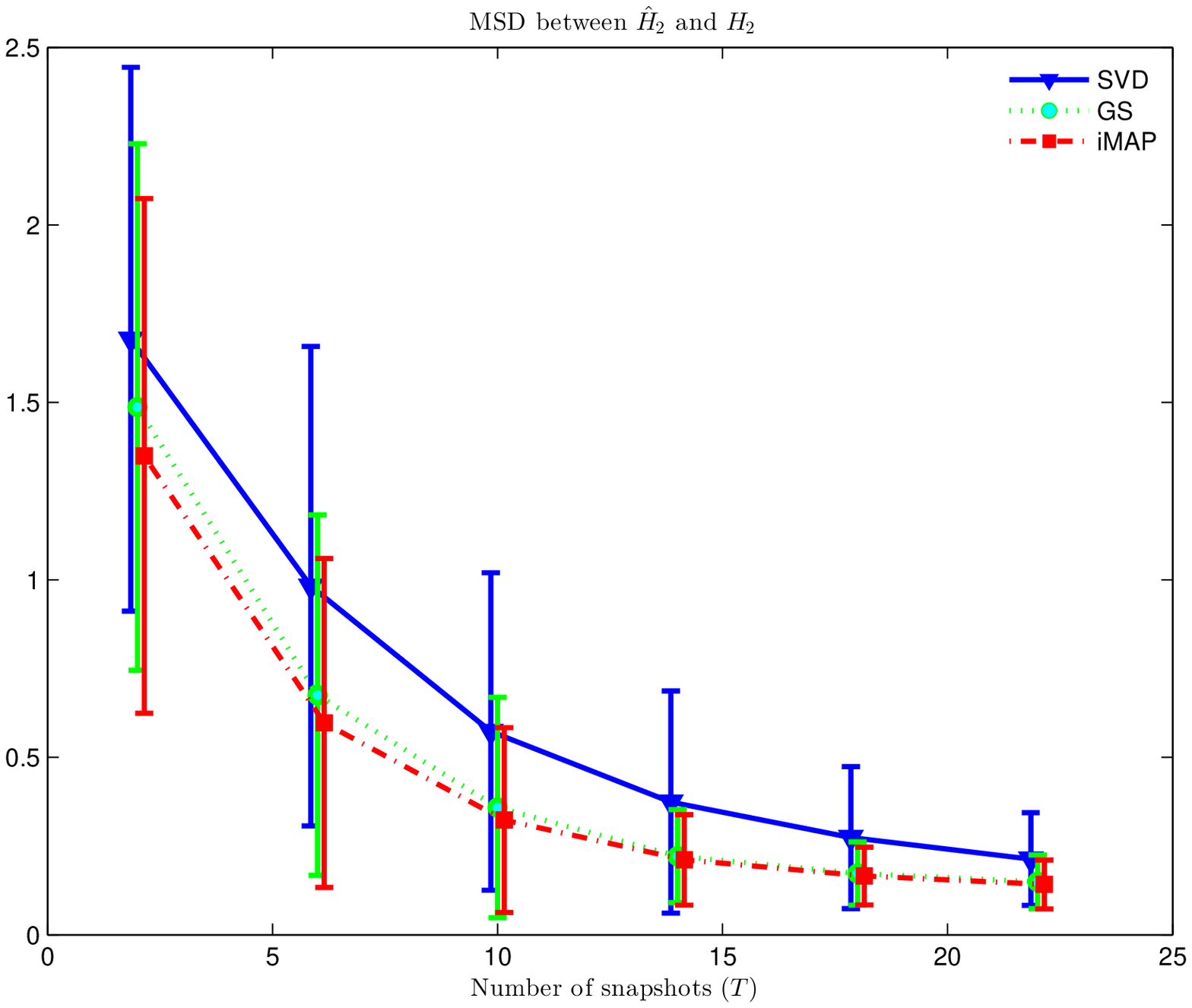}} \\
\subfigure[][]{\label{fig: theta1_vs_T_M=8_R=2_kappa=40_SNR=0}
\includegraphics[width=6cm]{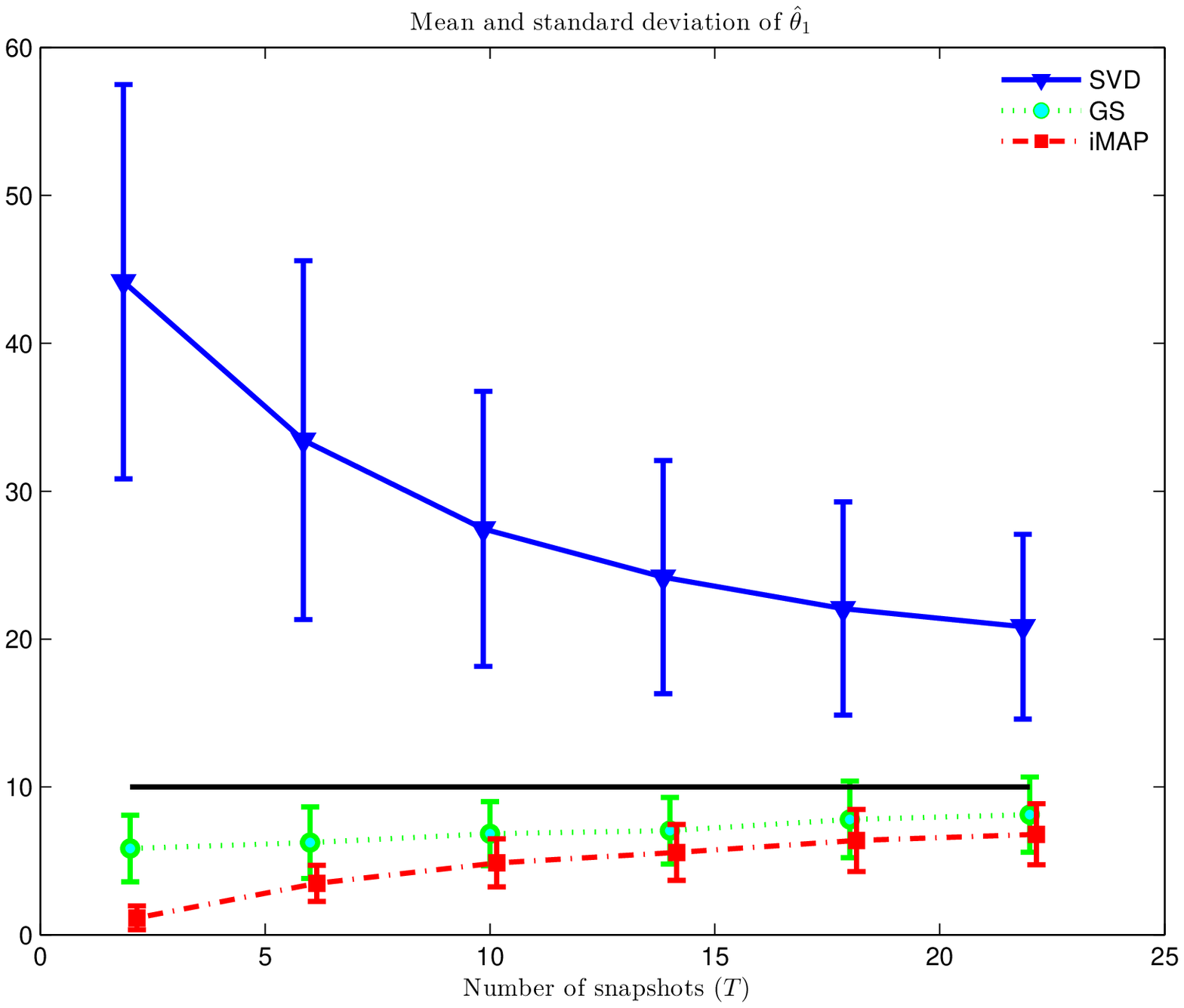}} \\
\subfigure[][]{\label{fig: theta2_vs_T_M=8_R=2_kappa=40_SNR=0}
\includegraphics[width=6cm]{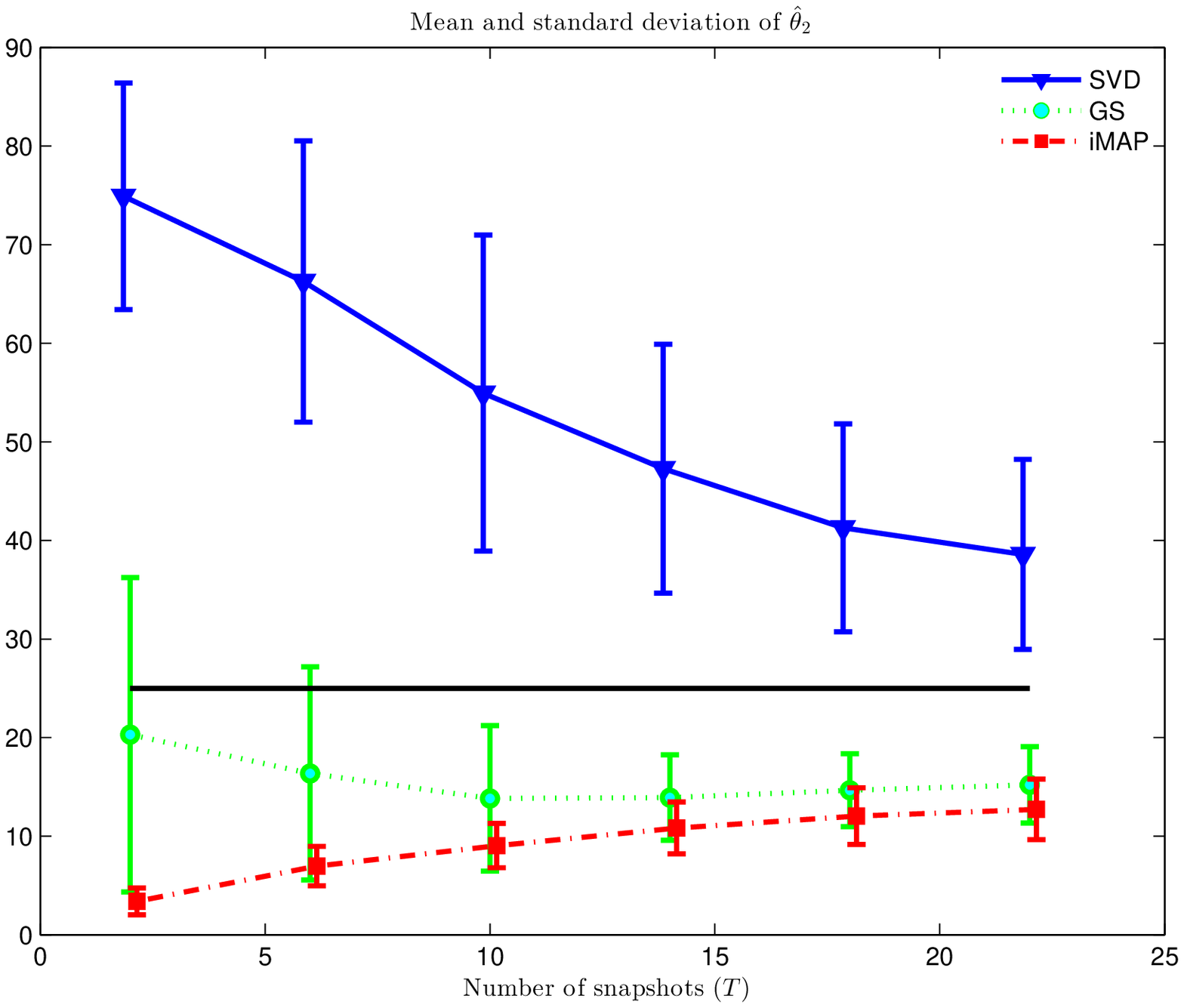}}
\caption{Performance of the estimators versus $T$. $\kappa=40$ and $\mathrm{SNR}=0$dB. \subref{fig: dist1_vs_T_M=8_R=2_kappa=40_SNR=0} $MSD(\Hhat_{1},\H_{1})$, \subref{fig: dist2_vs_T_M=8_R=2_kappa=40_SNR=0} $MSD(\Hhat_{2},\H_{2})$, \subref{fig: theta1_vs_T_M=8_R=2_kappa=40_SNR=0}, mean and std of $\hat{\theta}_{1}$,  \subref{fig: theta2_vs_T_M=8_R=2_kappa=40_SNR=0}, mean and std of $\hat{\theta}_{2}$.}
\label{fig: res_vs_T_M=8_R=2_kappa=40_SNR=0}
\end{figure}
\fi

\ifCLASSOPTIONdraftcls
\begin{figure}[p]
\centering
\subfigure[][]{\label{fig: dist1_vs_SNR_M=8_R=2_kappa=40_T=6}
\includegraphics[width=7cm]{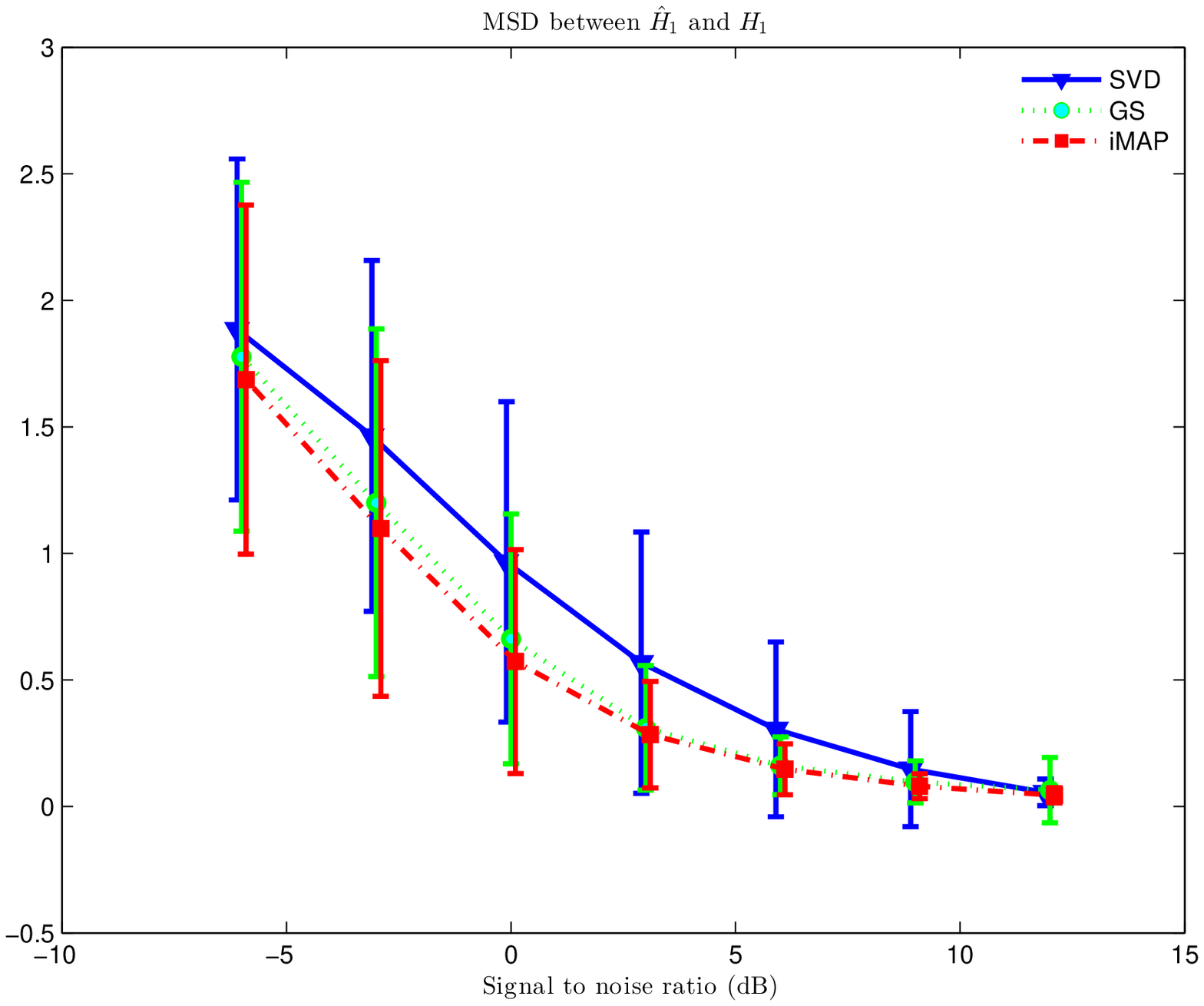}}
\subfigure[][]{\label{fig: dist2_vs_SNR_M=8_R=2_kappa=40_T=6}
\includegraphics[width=7cm]{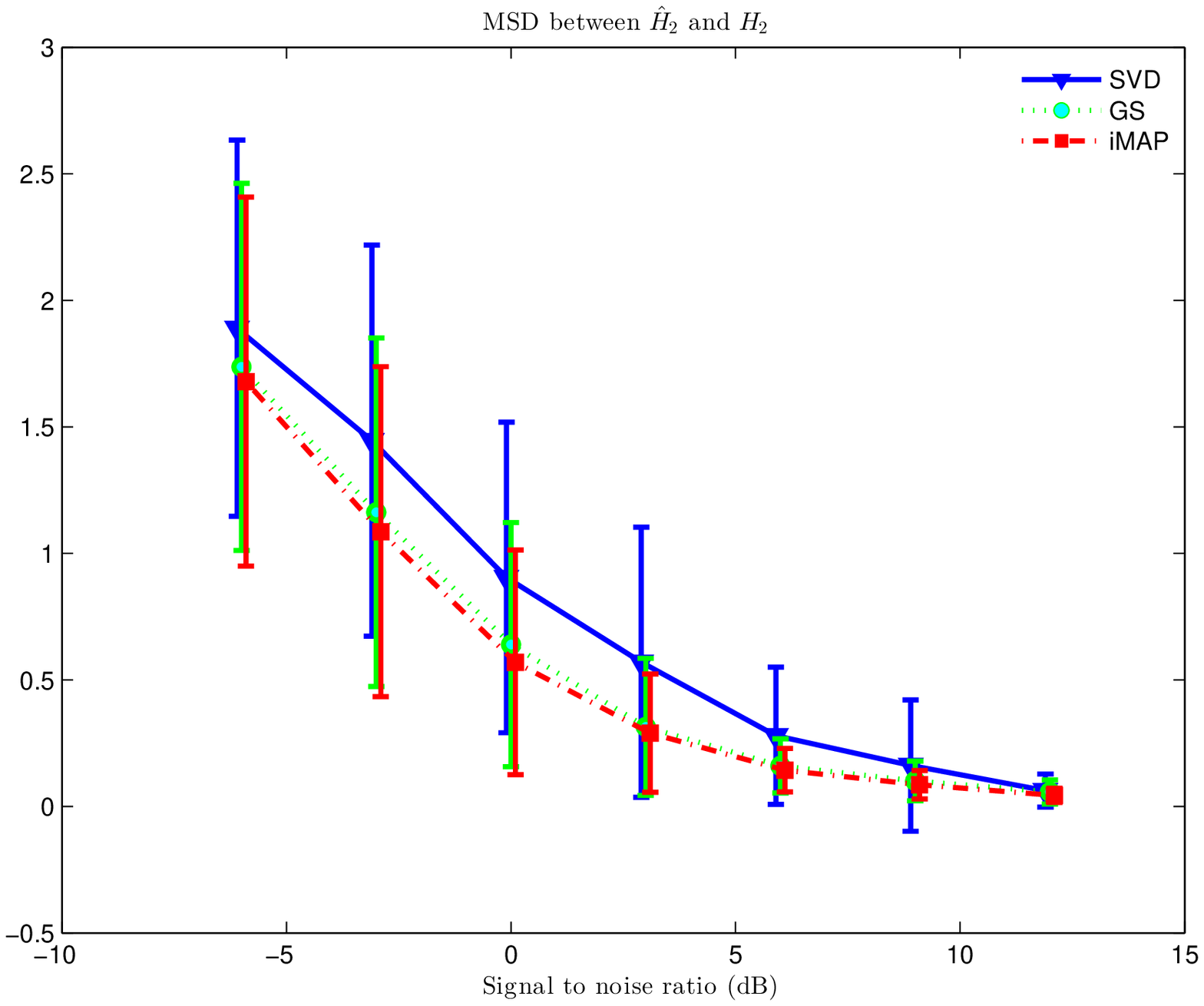}}\\
\subfigure[][]{\label{fig: theta1_vs_SNR_M=8_R=2_kappa=40_T=6}
\includegraphics[width=7cm]{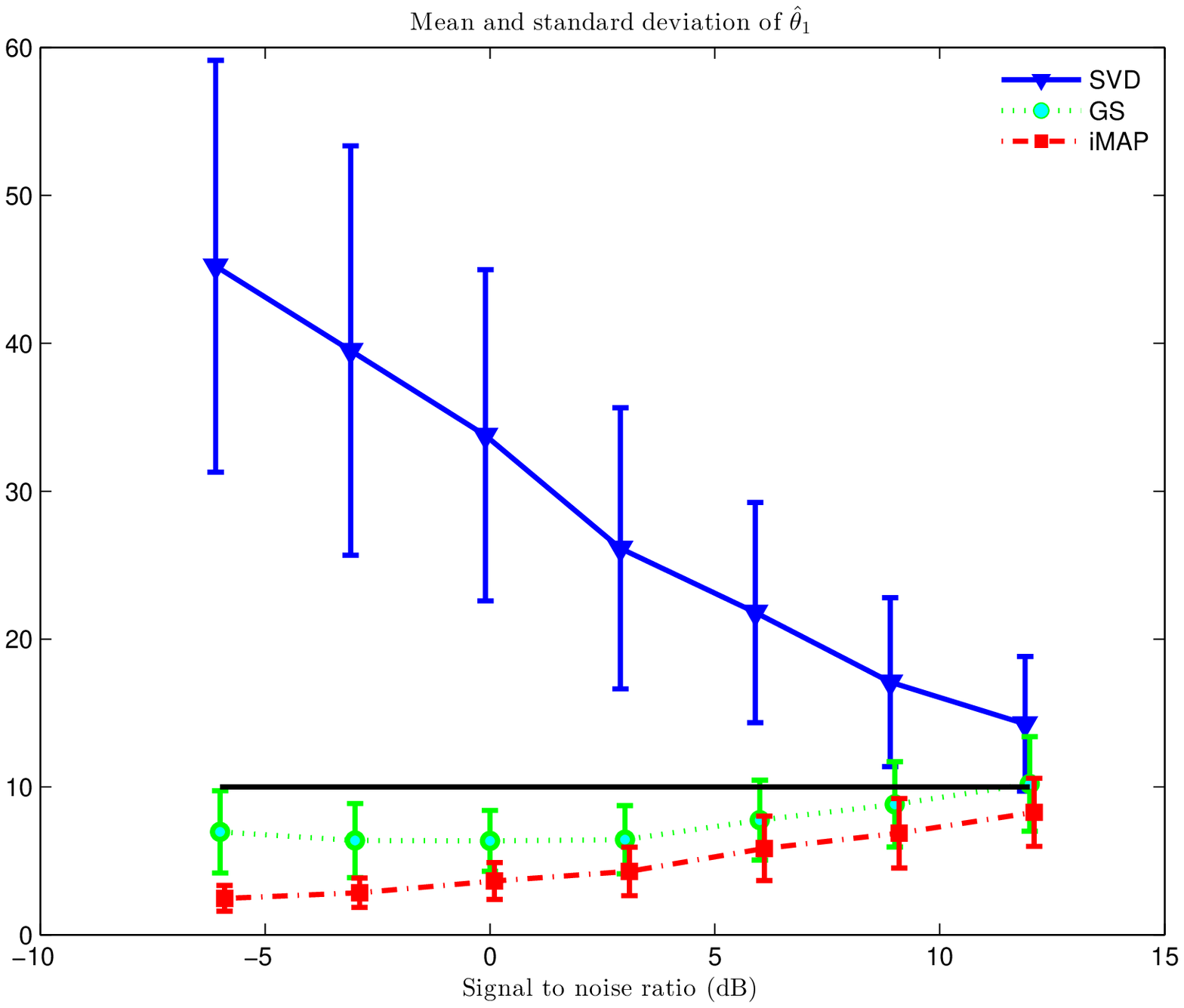}}
\subfigure[][]{\label{fig: theta2_vs_SNR_M=8_R=2_kappa=40_T=6}
\includegraphics[width=7cm]{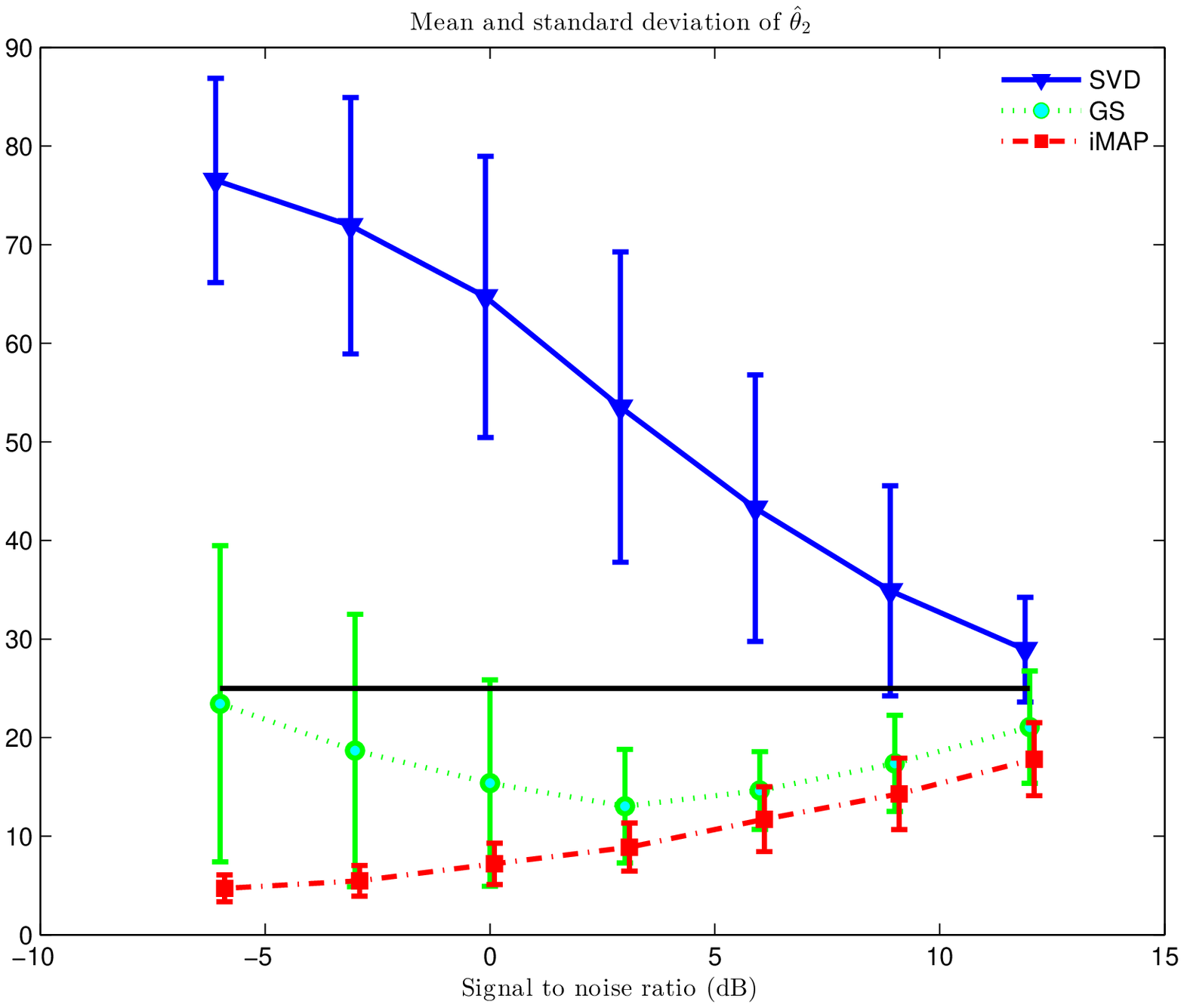}}
\caption{Performance of the estimators versus $\mathrm{SNR}$. $\kappa=40$ and $T=6$. \subref{fig: dist1_vs_SNR_M=8_R=2_kappa=40_T=6} $MSD(\Hhat_{1},\H_{1})$, \subref{fig: dist2_vs_SNR_M=8_R=2_kappa=40_T=6} $MSD(\Hhat_{2},\H_{2})$, \subref{fig: theta1_vs_SNR_M=8_R=2_kappa=40_T=6}, mean and std of $\hat{\theta}_{1}$,  \subref{fig: theta2_vs_SNR_M=8_R=2_kappa=40_T=6}, mean and std of $\hat{\theta}_{2}$.}
\label{fig: res_vs_SNR_M=8_R=2_kappa=40_T=6}
\end{figure}
\else
\begin{figure}[!t]
\centering
\subfigure[][]{\label{fig: dist1_vs_SNR_M=8_R=2_kappa=40_T=6}
\includegraphics[width=6cm]{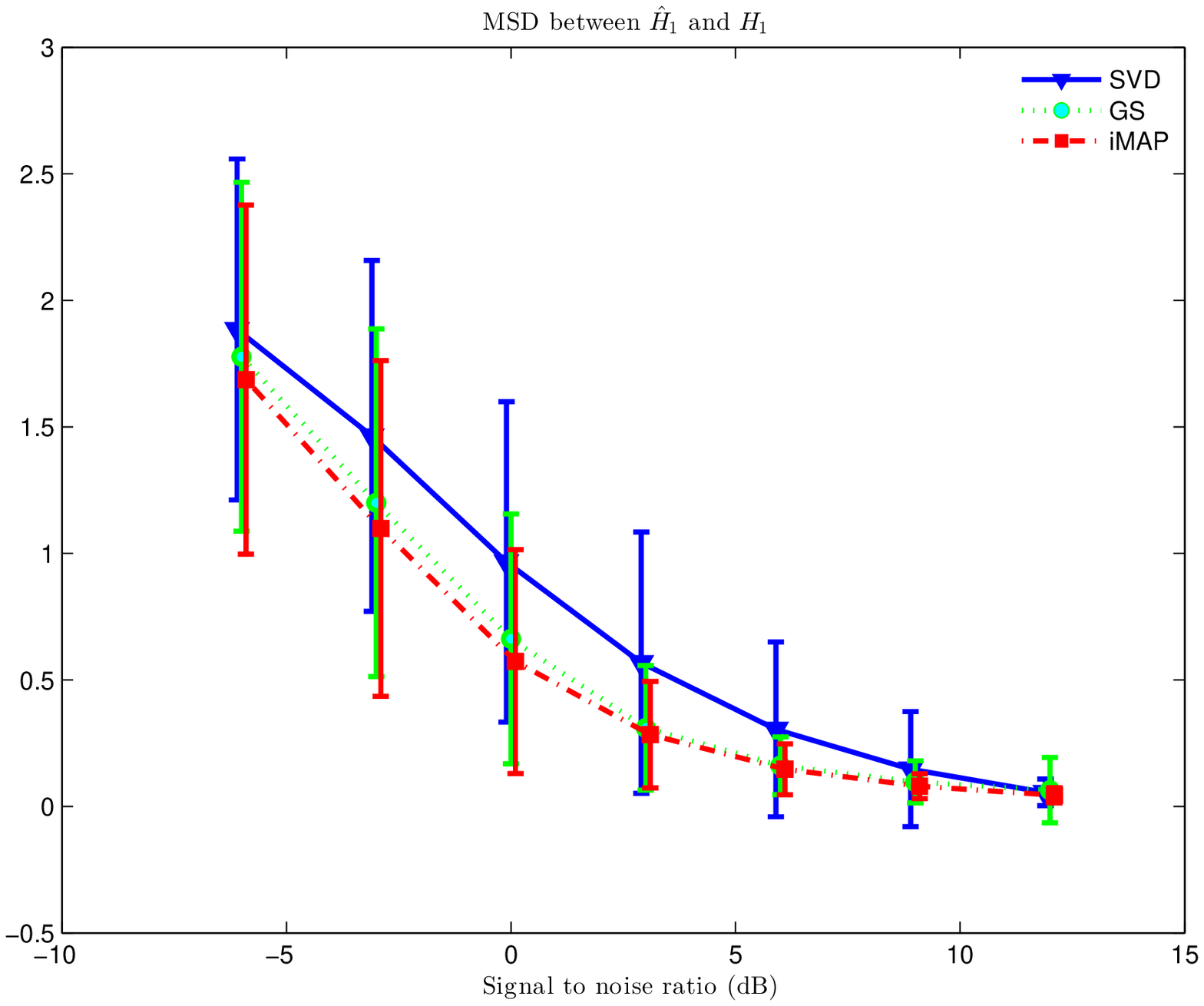}} \\
\subfigure[][]{\label{fig: dist2_vs_SNR_M=8_R=2_kappa=40_T=6}
\includegraphics[width=6cm]{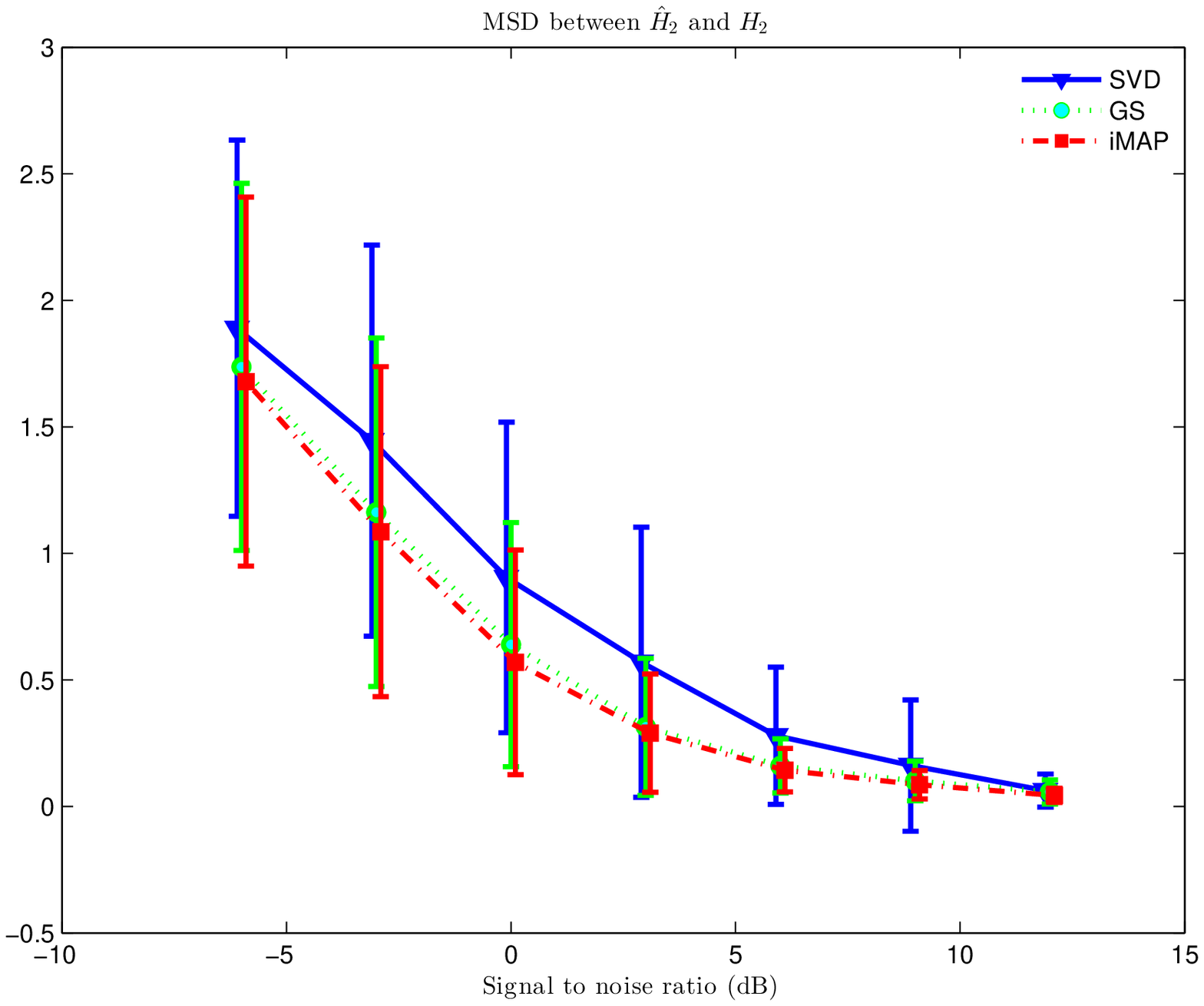}}\\
\subfigure[][]{\label{fig: theta1_vs_SNR_M=8_R=2_kappa=40_T=6}
\includegraphics[width=6cm]{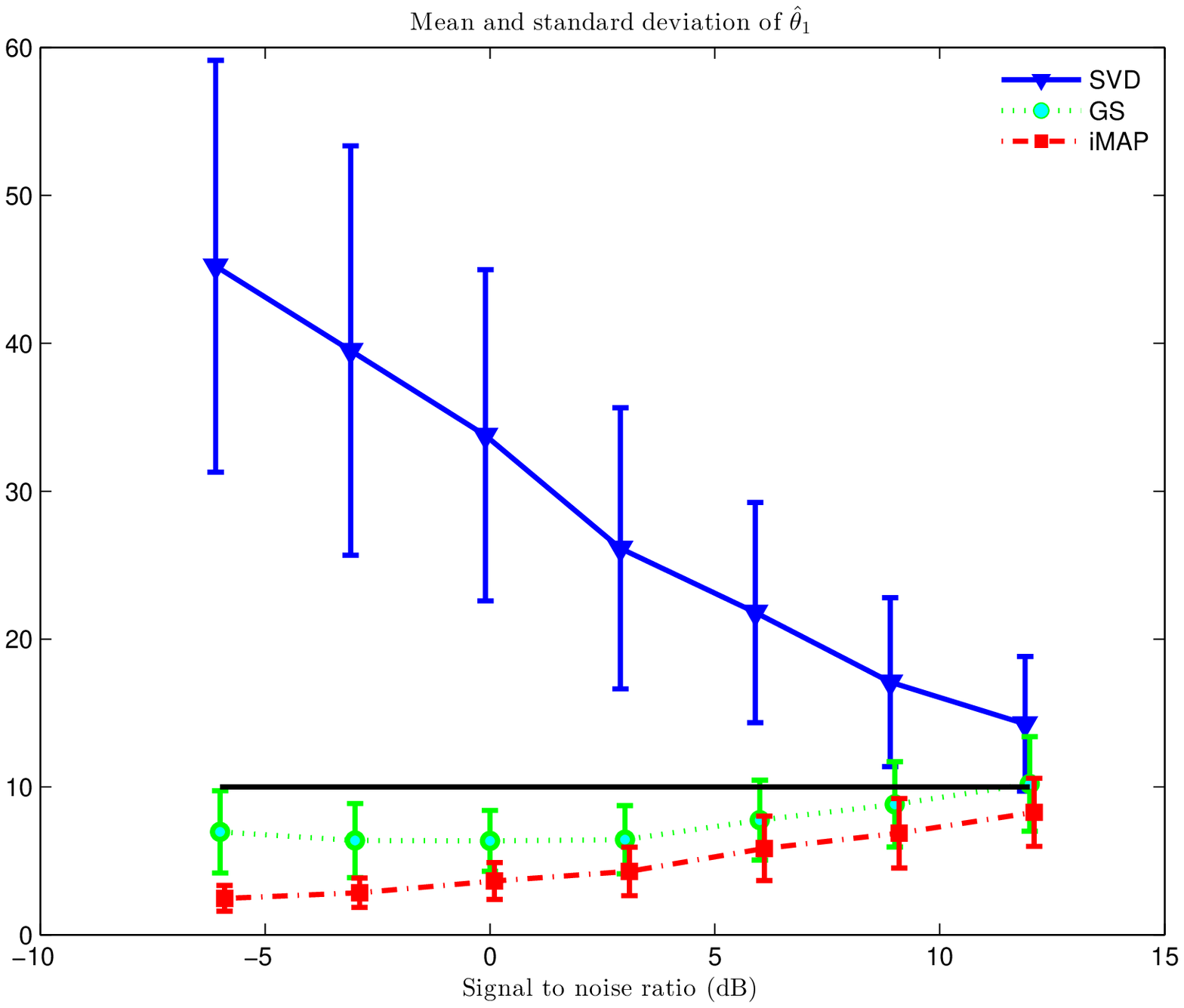}} \\
\subfigure[][]{\label{fig: theta2_vs_SNR_M=8_R=2_kappa=40_T=6}
\includegraphics[width=6cm]{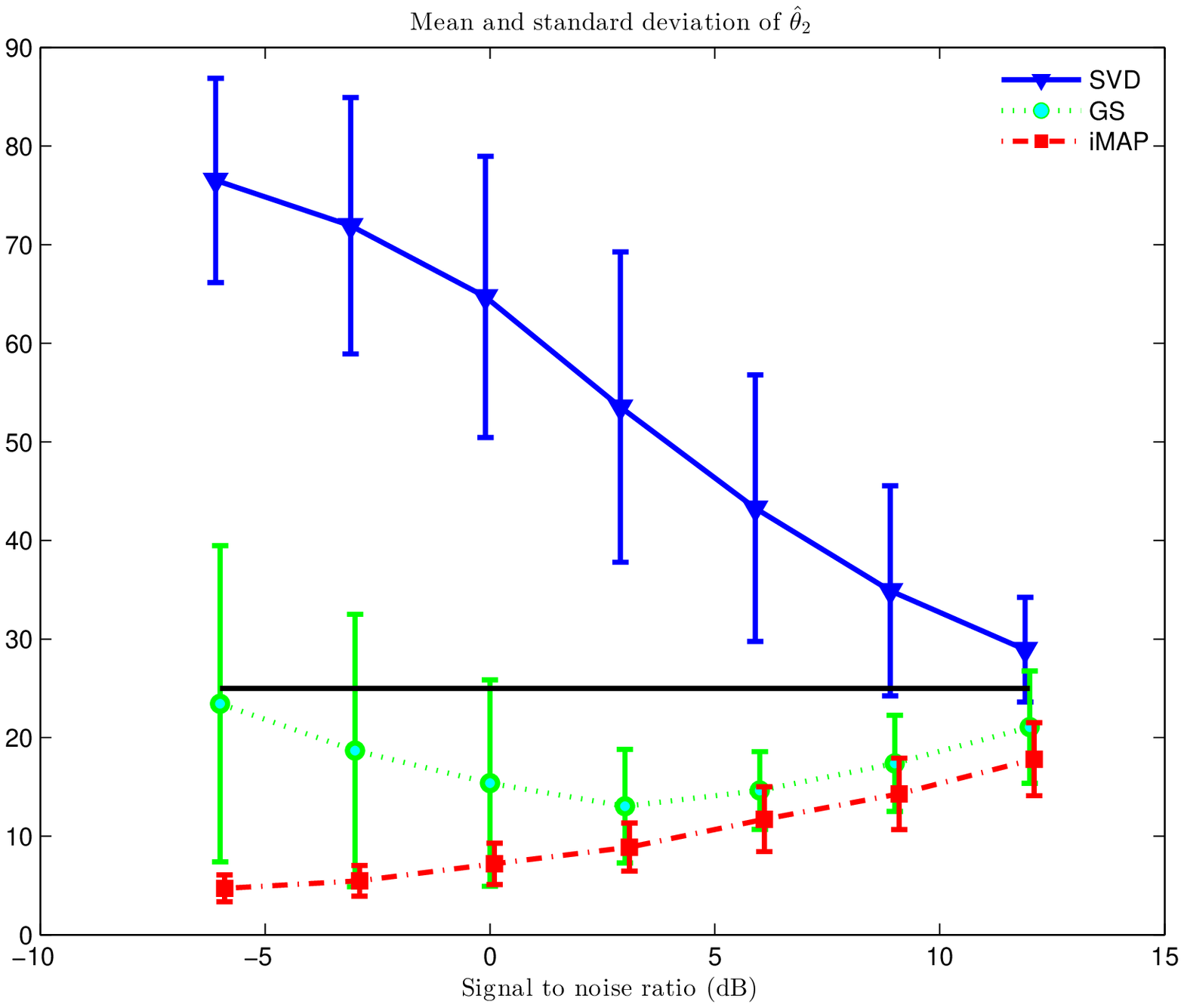}}
\caption{Performance of the estimators versus $\mathrm{SNR}$. $\kappa=40$ and $T=6$. \subref{fig: dist1_vs_SNR_M=8_R=2_kappa=40_T=6} $MSD(\Hhat_{1},\H_{1})$, \subref{fig: dist2_vs_SNR_M=8_R=2_kappa=40_T=6} $MSD(\Hhat_{2},\H_{2})$, \subref{fig: theta1_vs_SNR_M=8_R=2_kappa=40_T=6}, mean and std of $\hat{\theta}_{1}$,  \subref{fig: theta2_vs_SNR_M=8_R=2_kappa=40_T=6}, mean and std of $\hat{\theta}_{2}$.}
\label{fig: res_vs_SNR_M=8_R=2_kappa=40_T=6}
\end{figure}
\fi

\newpage
\bibliographystyle{IEEEtran}
\bibliography{bprocrus}
\end{document}